\documentclass[useAMS,usenatbib]{mn2e}
\usepackage{graphicx}
\title{Gravitational scattering of the sterile neutrino halo dark matter}
\author[ Chan ]{ Man Ho Chan \thanks{chanmh@eduhk.hk}
\\ Department of Science and Environmental Studies, The Education University of Hong Kong, Tai Po, Hong Kong}

\begin{document}

\date{Accepted XXXX, Received XXXX}

\pagerange{\pageref{firstpage}--\pageref{lastpage}} \pubyear{XXXX}

\maketitle

\label{firstpage}

\date{\today}

\begin{abstract}
A recent study shows that gravitational scattering of dark matter, in the form of massive objects with mass $m \sim 10^3-10^4M_{\odot}$, could provide a possible solution to alleviate the small-scale structure problems of cold dark matter. The scattering cross section is velocity-dependent so that this scenario can explain why self-interaction of dark matter is significant in dwarf galaxies, but not in massive galaxies and galaxy clusters. In this Letter, we show that this kind of dark massive objects could be made of sterile neutrinos with a possible rest mass range $m_{\nu} \sim 7.6$ keV $-$ 71 MeV. This mass range generally satisfies most of the current observational constraints. The entire structure of the sterile neutrino halos can be simply predicted from standard physics.
\end{abstract}


\section{Introduction}
The idea of self-interacting dark matter (SIDM) has been proposed for more than two decades. The main reason for introducing SIDM is to solve the small-scale structure problems of cold dark matter (CDM). The CDM model predicts that the central cold dark matter density exhibits a cusp-like profile \citep{Navarro}. This is generally true for massive galaxies \citep{Iocco,Sofue} and galaxy clusters \citep{Viola} while observations indicate that many dwarf galaxies exhibit core-like density profiles \citep{Zackrisson,deBlok}. The discrepancy between the CDM prediction and the core-like structures observed in dwarf galaxies is now known as the core-cusp problem \citep{deBlok}. In view of this problem, \citet{Spergel,Firmani} show that dark matter with self-interaction could modify the central density cusp to core. This effect has been verified by many recent numerical simulations \citep{Sameie,Silverman}. Moreover, the SIDM model can also help solve the too-big-to-fail problem \citep{Vogelsberger,Kaplinghat}. Therefore, the SIDM model has become a popular scenario to solve the small-scale structure problems in the CDM model.

To solve the small-scale structure problems, the self-interacting cross section per unit mass should be $\sigma/m \sim 1-10$ cm$^2$/g \citep{Kaplinghat,Silverman}. However, recent observations have placed stringent constraints on the upper limit of the self-interacting cross section per unit mass. For example, The study in \citet{Randall} analyze the data of the Bullet Cluster and obtain a strong upper limit $\sigma/m \le 1.25$ cm$^2$/g (68\% confidence). Recent analyses have improved the limit to $\sigma/m \le 0.35$ cm$^2$/g \citep{Peter,Sagunski}. Besides, there is no apparent discrepancy with the CDM model for massive galaxies and galaxy clusters. Therefore, \citet{Loeb} propose that the self-interacting cross section might be velocity-dependent so that self-interaction of dark matter is important in dwarf galaxies only (with low velocity). In particular, if the self-interaction is mediated by a Yukawa potential, the self-interacting cross section would be inversely proportional to the fourth power of dark matter velocity $v$ \citep{Loeb}. This idea can perfectly address the small-scale structure problems and maintain the success of the CDM model in large-scale structures simultaneously \citep{Loeb,Chan}.

Recently, a study has suggested a new scenario of velocity-dependent SIDM \citep{Loeb2}. By assuming that dark matter exists in the form of massive halos, the gravitational scattering among the massive halos can provide the required self-interaction to solve the small-scale structure problems. The term `self-interaction' here refers to the gravitational scattering of the massive halos. The cross section per unit mass has a velocity dependence of $v^{-4}$ and the possible mass range of the halos is $m \sim 10^3-10^4M_{\odot}$ \citep{Loeb2}. In this Letter, by following standard quantum and gravitational physics, we show that sterile neutrinos can form this kind of dark massive halos. The possible range of sterile neutrino mass is $\sim 7.6$ keV $-$ 71 MeV, which is consistent with the proposals of the cosmological warm dark matter \citep{Dodelson,Shi}.

\section{Sterile neutrino halos}
Observations indicate that active neutrinos have rest mass \citep{Fukuda}, which possibly suggests that right-handed neutrinos might also exist. Many particle physics models propose that there exist the 4th-type neutrinos called sterile neutrinos with mass larger than keV \citep{Adhikari}. They can be produced via resonant mechanism (active-sterile neutrino conversion) \citep{Shi} or non-resonant mechanism (oscillation between active and sterile neutrinos) \citep{Dodelson} in the early universe. They are commonly proposed as a candidate of warm dark matter. Many recent studies have focused on the decaying properties of sterile neutrinos \citep{Chan2,Chan3,Bulbul,Boyarsky}.

Sterile neutrinos are fermions. They have non-zero rest mass and they would collapse gravitationally. However, this gravitational collapse would not necessarily form a black hole because sterile neutrinos would exert quantum degeneracy pressure $P(r)$ when their density $\rho_{\nu}(r)$ is high:
\begin{equation}
P(r)=\frac{h^2}{5m_{\nu}^{8/3}} \left(\frac{3}{4\pi g_s} \right)^{2/3} [\rho_{\nu}(r)]^{5/3}=K[\rho_{\nu}(r)]^{5/3},
\end{equation}
where $g_s$ is the particle spin degeneracy of sterile neutrinos. In the followings, we assume $g_s=2$ for simplicity \citep{Gomez}. This quantum degeneracy pressure can balance the gravitational attraction of the sterile neutrinos to achieve a hydrostatic equilibrium:
\begin{equation}
\frac{dP(r)}{dr}=-\frac{Gm(r)\rho_{\nu}(r)}{r^2},
\end{equation}
where
\begin{equation}
m(r)=4 \pi \int_0^r r'^2\rho_{\nu}(r')dr'.
\end{equation}
We can combine the above three equations to form the Lane-Emden equation with the polytropic index $n=3/2$ \citep{Domcke,Gomez}: 
\begin{equation}
\frac{1}{\xi^2}\frac{d}{d \xi} \left(\xi^2 \frac{d\theta}{d\xi} \right)=-[\theta(\xi)]^{3/2},
\end{equation}
where $\rho_{\nu}(\xi)=\rho_c [\theta(\xi)]^{3/2}$, $r=\xi (5K\rho_c^{-1/3}/8 \pi G)^{1/2}$, and $\rho_c$ is the central density.

Solving the Lane-Emden equation, the radius $R$ and the total mass $m$ of the sterile neutrino halo are \citep{Domcke,Gomez}
\begin{equation}
R=3.654 \left(\frac{5K}{8\pi G} \right)^{1/2} \rho_c^{-1/6},
\end{equation}
and 
\begin{equation}
m=34.11 \left(\frac{5K}{8 \pi G} \right)^{3/2} \rho_c^{1/2}
\end{equation}
respectively. By combining Eqs.~(5) and (6), we can get the relation between $R$ and $m$ \citep{Domcke}:
\begin{equation}
R=11.85\left( \frac{5K}{8 \pi G} \right)m^{-1/3}=192~{\rm pc} \left(\frac{m_{\nu}}{1~\rm keV} \right)^{-8/3} \left(\frac{m}{10^4M_{\odot}} \right)^{-1/3}.
\end{equation}

Generally speaking, the time required for the gravitational collapse of sterile neutrinos can be estimated by the free-falling time \citep{Phillips}:
\begin{equation}
t_{ff}=\sqrt{\frac{3\pi}{32G \langle \rho(z) \rangle}},
\end{equation}
where $\langle \rho(z) \rangle$ is the average cosmological dark matter density at redshift $z$. As discussed in \citet{Loeb2}, the minimum formation redshift of the dark matter halo could be $z>700$. If the sterile neutrinos begin to form the halos at $z=700$, we have $\langle \rho(z) \rangle \sim 1.4 \times 10^{10}M_{\odot}$ kpc$^{-3}$ and $t_{ff} \sim 2$ Myr ($\sim 7 \times 10^{13}$ s). Based on this estimated time $t_{ff} \sim 2$ Myr, the formation of the halos would be finished before $z \sim 200$, which satisfies the criterion of the first virialized mini-halo formation at $z \sim 70$ \citep{Loeb2}. In fact, similar studies have been done previously in examining the behaviors of the Fermion ball dark matter \citep{Munyaneza,Domcke,Gomez} or neutrino halo dark matter \citep{Viollier,Chan2}. However, the scopes of these studies are entirely different from that in this study.

\section{Constraints of the sterile neutrino mass}
According to the gravitational scattering model proposed by \citet{Loeb2}, the physical size of the dark massive halo must be smaller than the minimum impact parameter $b_{\rm min}=2Gm/v^2$ for the gravitational scattering. Therefore, we have $R<2Gm/v^2$. The first constraint is thus given by:
\begin{equation}
223\left(\frac{m_{\nu}}{1~{\rm keV}} \right)^{-8/3} \left(\frac{v}{10~{\rm km/s}} \right)^2< \left(\frac{m}{10^4M_{\odot}} \right)^{4/3}.
\end{equation}
Generally speaking, a small value of $m_{\nu}$ would give a large $R$. Therefore, the constraint in Eq.~(9) sets the lower bound of the sterile neutrino mass $m_{\nu}$.

On the other hand, if the sterile neutrino density is very high, the degenerate sterile neutrinos are relativistic and the equation of state would change to $P(r) \propto [\rho_{\nu}(r)]^{4/3}$. In this case, the situation would be similar to the case of a white dwarf \citep{Phillips}. There exists an upper limit of the mass (the Chandrasekhar mass) for the sterile neutrino halos. Therefore, we can write the `Chandrasekhar limit' of the sterile neutrino halos as
\begin{equation}
M_{\rm Ch} \approx 5.0 \times 10^{12}M_{\odot} \left(\frac{m_{\nu}}{1~\rm keV} \right)^{-2}.
\end{equation}
Since the value of $m$ must be smaller than the Chandrasekhar limit (i.e. $m<M_{\rm Ch}$), we can get the second constraint:
\begin{equation}
m < 5.0 \times 10^{12}M_{\odot} \left(\frac{m_{\nu}}{1~\rm keV} \right)^{-2}.
\end{equation}
Here, as a large value of $m_{\nu}$ would give a small $M_{\rm Ch}$, the constraint in Eq.~(11) sets the upper bound of the sterile neutrino mass $m_{\nu}$. Note that the Chandrasekhar limit here is the mass limit of the case with the largest possible mass density of sterile neutrinos (infinitely large), which is equivalent to binding the relativistic sterile neutrinos together. For binding non-relativistic sterile neutrinos, the maximum possible mass of the halo must be smaller than the Chandrasekhar limit. Nevertheless, it is very difficult to determine the robust upper mass limit for non-relativistic sterile neutrinos as the actual maximum central mass density is uncertain. Therefore, the Chandrasekhar limit used here is a very conservative limit.

As pointed out in \citet{Loeb2}, the dark matter self-interacting (gravitational scattering) cross section per unit mass is given by
\begin{equation}
\frac{\sigma}{m}=10~{\rm cm^2/g} \left[\frac{(m/10^4M_{\odot})}{(v/10~{\rm km/s})^4}\right].
\end{equation}
To solve the core-cusp problem in dwarf galaxies, the expected cross section per unit mass is $\sigma/m \sim 1-10$ cm$^2$/g \citep{Kaplinghat,Silverman}. Assuming the fiducial value of the velocity in dwarf galaxies $v \sim 10$ km/s \citep{Loeb2}, the possible range of $m$ should be $10^3-10^4M_{\odot}$. Note that the value of $\sigma/m$ is proportional to $mv^{-4}$. A smaller value of $v$ would give a smaller possible upper limit of $m$. Nevertheless, some values of $v$ obtained by observational data are smaller than $v=10$ km/s (e.g., Carina dwarf galaxy) \citep{Walker}. Therefore, the upper limit of $m \sim 10^4M_{\odot}$ considered in this study based on the fiducial value $v=10$ km/s is somewhat conservative. By considering the range $m=10^3-10^4M_{\odot}$ and the two constraints in Eqs.~(9) and (11), the possible range of $m_{\nu}$ is $\sim 7.6$ keV $-$ 71 MeV, assuming $v=10$ km/s (see Fig.~1). 

Moreover, we can also estimate the possible range of $m_{\nu}$ by using the Jeans mass analysis. The sterile neutrinos would undergo gravitational collapse when the mass of halo $m$ is larger than the Jeans mass $M_J$. The Jeans mass is given by:
\begin{equation}
M_J \approx \frac{\pi c_s^3}{6G^{3/2} \langle \rho_{\nu} \rangle^{1/2}},
\end{equation}
where $\langle \rho_{\nu} \rangle$ is the average sterile neutrino density at the formation time and $c_s=\sqrt{dP/d\rho_{\nu}}=\sqrt{(5/3)K \langle \rho_{\nu} \rangle^{2/3}}$ is the sound speed. If we take the typical value of the central density of a dwarf galaxy as the average sterile neutrino density $\langle \rho_{\nu} \rangle=3\times 10^7M_{\odot}$ kpc$^{-3}$ for an estimation \citep{Loeb2}, we get
\begin{equation}
M_J=1.4\times 10^4M_{\odot} \left(\frac{m_{\nu}}{1~\rm keV} \right)^{-4}.
\end{equation}
Therefore, for $M_J=10^3M_{\odot}-10^4M_{\odot}$, we get $m_{\nu} \sim 1-2$ keV. In other words, for $m_{\nu}<1$ keV, the halo mass with $m<10^4M_{\odot}$ would be difficult to form. Therefore, our proposed range of $m_{\nu}$ can satisfy the Jeans criterion for the gravitational collapse of sterile neutrino halos. Generally speaking, there is no upper limit of $m_{\nu}$ based on the Jeans mass analysis.

\section{Discussion}
In this Letter, we have investigated the possibility of the degenerate sterile neutrino halos being the self-interacting dark massive objects suggested in \citet{Loeb2}. Note that sterile neutrinos do not have self-interaction except gravity. The term `self-interacting' here refers to the gravitational scattering of the massive objects formed by the sterile neutrinos. The sterile neutrinos formed in the early universe would collapse into the dark massive objects due to self-gravitational attractive force. The required formation time is much shorter than the current cosmological age. By using the known properties of neutrinos (can exert quantum degeneracy pressure) and following standard gravitational physics, we can get the intrinsic relation of $m$ and $R$ and constrain the allowed range of $m_{\nu}$. The sterile neutrino halos can be viewed as a large dark matter particle (with size $<1$ pc) and they behave like cold dark matter. Moreover, the gravitational scattering of these dark massive objects can help explain the small-scale problem in dwarf galaxies because the cross section is strongly velocity-dependent. Here, we have shown that a wide range of sterile neutrino mass ($\sim 7.6$ keV $-$ 71 MeV) can form the suggested dark massive objects. This range of $m_{\nu}$ can also satisfy the Jeans criterion for the formation of $\sim 10^3-10^4M_{\odot}$ sterile neutrino halos at the centre of a dwarf galaxy.

The theoretical model-independent lower limit of sterile neutrino mass can be constrained by the Tremaine-Gunn bound \citep{Tremaine}. Using current data, the Tremaine-Gunn lower bound for fermionic dark matter is about $\sim 100$ eV \citep{Davoudiasl}. For the specific non-resonant production mechanism for sterile neutrinos, the most stringent Tremaine-Gunn lower bound is $m_{\nu}>2.79$ keV \citep{Boyarsky2}. Therefore, our constrained range is far above these lower bounds. Moreover, if sterile neutrino can decay, the sterile neutrino mass $m_{\nu}$ can be constrained by the X-ray/gamma-ray flux limit observations. However, since the decay rate depends on the model-dependent mixing angle $\theta$ \citep{Barger}, the X-ray/gamma-ray flux limits can only constrain a certain parameter space of $\theta$ and $m_{\nu}$ \citep{Boyarsky2,Roach}, unless there is a sharp excess line observed. For some specific production mechanisms, the mixing angle determines both the sterile neutrino dark matter abundance and decay rate so that the parameter space of $\theta$ and $m_{\nu}$ can be further constrained \citep{Boyarsky2,Ng,Roach}. In these cases, X-ray/gamma-ray flux limits can provide constraints for $m_{\nu}$. Nevertheless, the most stringent constraints based on the X-ray data of the M31 galaxy and our Galaxy still allow $m_{\nu} \ge 7$ keV \citep{Ng,Roach}. Generally speaking, our proposed range of $m_{\nu}$ can satisfy most of the current bounds \citep{Gelmini}. 

If our model is correct, the dark matter halo mass $m$ is not constant because it depends on the central mass density $\rho_c$. Therefore, there might exist a distribution of $m$ inside a galaxy or a galaxy cluster. Moreover, it is also possible that some sterile neutrinos are completely free (without forming any massive halos). If the amount of the free sterile neutrinos is small, then it would be very difficult to get any positive signals from the direct-detection experiments of sterile neutrinos \citep{Campos,Shoemaker}.

Some previous studies have claimed the discoveries of the decaying sterile neutrino signals with $m_{\nu} \approx 7$ keV \citep{Bulbul,Boyarsky} or $m_{\nu} \approx 17$ keV \citep{Prokhorov,Chan3}. However, these discoveries are quite controversial now \citep{Bhargava,Dekker,Silich}. In particular, the claimed value of $m_{\nu} \approx 7$ keV is just marginally consistent with our possible range obtained. This value has also been challenged by the recent studies of the Lyman-$\alpha$ forest \citep{Garzilli,Enzi}, strong lensing \citep{Vegetti,Enzi}, satellite galaxy count \citep{Cherry,Nadler}, and the 21-cm signal \citep{Vipp}. In particular, the 21-cm data have placed a lower limit $m_{\nu} \ge 63^{+19}_{-35}$ keV for the non-resonant production mechanism \citep{Vipp}. Our results combining with the other recent bounds generally favor $m_{\nu} \ge 10$ keV. In view of this, the decay signals of sterile neutrinos might provide a direct evidence of the existence of sterile neutrinos. These potential signals could be observed by future X-ray or MeV gamma-ray telescopes.

\begin{figure}
\vskip 10mm
 \includegraphics[width=90mm]{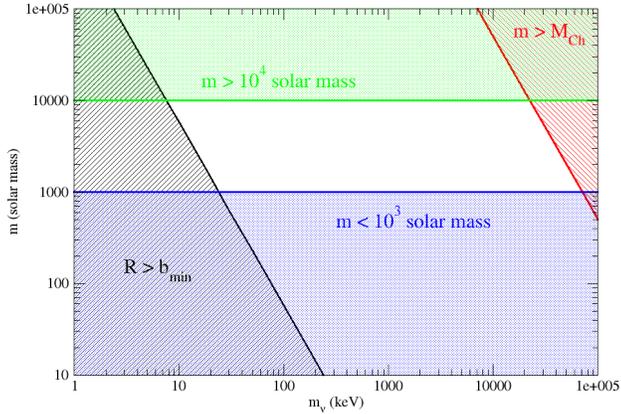}
 \caption{The unshaded region represents the allowed parameter space for $m_{\nu}$ and $m$. The shaded regions are the forbidden regions constrained by the conditions $R<b_{\rm min}$, $m<M_{\rm Ch}$ and $10^3M_{\odot}<m<10^4M_{\odot}$. Here, we have assumed $v=10$ km/s.}
\vskip 10mm
\end{figure}

\section{Acknowledgements}
We thank the anonymous referee for useful constructive feedback and comments. The work described in this paper was partially supported by the Seed Funding Grant (RG 68/2020-2021R) and the Dean's Research Fund of the Faculty of Liberal Arts and Social Sciences, The Education University of Hong Kong, Hong Kong Special Administrative Region, China (Project No.: FLASS/DRF 04628).

\section{Data availability statement}
The data underlying this article will be shared on reasonable request to the corresponding author.

\label{lastpage}

\end{document}